# Retrieving refractive index of single spheres using the phase spectrum of light-scattering pattern

Andrey V. Romanov,[a,b] Valeri P. Maltsev,[a,b] and Maxim A. Yurkin[a,b*]

[a] *Voevodsky Institute of Chemical Kinetics and Combustion SB RAS, Institutskaya Str. 3, 630090 Novosibirsk, Russia*

[b] *Novosibirsk State University, Pirogova Str. 2, 630090 Novosibirsk, Russia*

*Corresponding author: yurkin@gmail.com

## Abstract

We analyzed the behavior of the complex Fourier spectrum of the angle-resolved light scattering pattern (LSP) of a sphere in the framework of the Wentzel–Kramers–Brillouin (WKB) approximation. Specifically, we showed that the phase value at the main peak of the amplitude spectrum almost quadratically depends on the particle refractive index, which was confirmed by numerical simulations using both the WKB approximation and the rigorous Lorenz–Mie theory. Based on these results, we constructed a method for characterizing polystyrene beads using the main peak position and the phase value at this point. We tested the method both on noisy synthetic LSPs and on the real data measured with the scanning flow cytometer. In both cases, the spectral method was consistent with the reference non-linear regression one. The former method leads to comparable errors in retrieved particle characteristics but is 300 times faster than the latter one. The only drawback of the spectral method is a limited operational range of particle characteristics that need to be set a priori due to phase wrapping. Thus, its main application niche is fast and precise characterization of spheres with small variation range of characteristics.

# 1   Introduction

Light scattering is a widely used research tool, exemplified by many measuring instruments for non-invasive studies of particles with sizes comparable to the wavelength. While a majority of instruments are based on the analysis of particle ensembles, the methods capable of measuring signals, especially light scattering patterns (LSPs), from single particles have greater reliability due to the underlying inverse problems being generally well-posed [1]. The latter methods usually assume a certain particle model and can be further divided into three broad classes, each having its pros and cons. The first and most common one is a non-linear regression. The solution of the direct light-scattering problem is fitted to the measured data, thus, accounting for all information contained in the LSP [2–5]. Although it provides, in many cases, the best accuracy and allows one to estimate the confidence interval of the obtained particle characteristics, it is relatively slow and its reliability is difficult to predict in advance.

The second class is based on the neural networks or, more generally, machine learning. This approach is fast after the training is complete but requires fine tuning. Its success in light-scattering characterization has been moderate so far [6–8], but it may further benefit from recent development in deep learning. The last class includes low-dimensional inversion methods. They compress all the information in a LSP down to a few parameters that determine the characteristics of the particle. These methods are mostly limited to simple particle models but are the fastest ones in this niche [9–11]. Moreover, they are amenable to detailed analysis, including that of robustness, through extensive numerical simulations.

Spectral methods are prominent examples of the latter class. Physically, they rely on the wave nature of the signal and the resulting interference (oscillations) observed in the LSP. Originally, such methods were built on the basis of the semi-empirical linear relationship between the main frequency contained in the signal (a peak in the amplitude Fourier spectrum of the LSP) and the particle size [12–16]. Later, other parameters of the amplitude spectrum were used to additionally determine the refractive index from the amplitude spectrum [17,18]. Moreover, the phase spectrum and other parameters of the amplitude spectrum were used to estimate or identify non-sphericity [19,20]. Recently, an important step towards the theoretical substantiation of all these semi-empirical methods has been made in Ref. [21], rigorously proving the relationship between the particle size and the main frequency of the spectrum. One of the implications of this theory is the relation between the refractive index and the LSP spectrum, including its phase.

The goal of this paper is to capitalize on this relation, by constructing a practical characterization method for spheres, which uses both the amplitude and phase spectrums. In section 2 we extend the theory in the framework of the Wentzel–Kramers–Brillouin (WKB) approximation to relate the particle refractive index with both the position and phase of the main spectral peak. Next, we rigorously calculate a set of LSPs in a relatively narrow range of sizes corresponding to 4 μm polystyrene beads in water (limited by phase wrapping) and construct an interpolant for the mapping from two spectral parameters into two sphere characteristics. In Section 3, we carry out an experimental verification of the developed method, using synthetic noisy data and real measurements of polystyrene beads with the scanning flow cytometer (SFC). Section 4 concludes the paper.

## 2 Spectral method

In the following we present a spectral method for characterization of a single sphere with size (diameter) $d$ and relative refractive index $m = n/n_0$ ($n$ is the particle refractive index, $n_0$ is the host medium index), illuminated by monochromatic plane wave with wavelength $\lambda$, using the measured angle-resolved light scattering profile (LSP). When no size units are specified, we use a dimensionless size $kd$ omitting $k$, where $k$ is the wavenumber given by $k = 2\pi n_0/\lambda$.

### 2.1 Measured signal and Fourier transformation

We work with a standard LSP measured by a SFC. It is a sum of $S_{11}$ and $S_{14}$ elements of the Muller scattering matrix [22], integrated over the azimuthal angle and considered at polar angles from $\theta_1 = 10°$ to $\theta_2 = 65°$:

$$I(\theta) = \frac{1}{2\pi} \int_0^{2\pi} [S_{11}(\theta, \varphi) + S_{14}(\theta, \varphi)] d\varphi, \tag{1}$$

where the integral of $S_{14}$ is exactly zero for any axisymmetric particle [23].

We apply the same spectral transformation as in [19] with the change of scattering coordinate to $\xi = 2\sin\theta/2$:

$$F(v) = \frac{1}{\xi_2 - \xi_1} \int_{\xi_1}^{\xi_2} w(\xi) I(\xi) \exp(-iv\xi) \, d\xi \tag{2}$$

where $w(\xi)$ is a Hann window function. Other window functions can also be used, for example, the Blackman-Nuttall window [18,20]. The numerical calculations are performed with the fast Fourier transform. In order to obtain the discrete spectrum as close as possible to the continuous one, we use following formula

$$F(v_k) = \frac{\exp\left(-i\xi_1 \frac{2\pi}{M} \frac{k}{\Delta\xi}\right)}{N} \sum_{j=0}^{N-1} w(\xi_1 + j\Delta\xi) I(\xi_1 + j\Delta\xi) \exp\left(-i\frac{2\pi}{M} kj\right), \tag{3}$$

where $\Delta\xi = (\xi_2 - \xi_1)/N$ and $v_k = 2\pi k/M\Delta\xi$, $N = 256$ is number of LSP sampling intervals and $M = 4096$ is number of resulting zero-padded discretization.

*2.2 Spectral parameters*

Previously, two parameters of the amplitude spectrum were used to characterize a sphere: the position of the main peak and the zero frequency amplitude, highly correlating with size and refractive index respectively [17]. Later these two parameters were tuned to obtain a solution in a wider range of characteristics [18]. The phase spectrum found applications in non-sphericity estimation [19] and classification [20]. Here we consider the use of the phase spectrum for characterization and start with the results obtained in Ref. [21].

Briefly, the amplitude of the zero peak corresponds to the LSP integral in the whole angular range of measurements, which is proportional to the refractive index in the WKB and Rayleigh-Debye-Gans (RGD) approximations. The main peak in the spectrum arises due to the features of this windowed Fourier transform. In the RGD approximation, the LSP is the Fourier transform into $\xi$ coordinate of the integrated (averaged in some directions) autocorrelation function (IAF) of the particle volume. The IAF can be considered as the intersection volume of the particle and its shifted copy, and it need to be replaced by the autocorrelation function of the relative refractive index in the case of an inhomogeneous particle. This function has a natural discontinuity of some derivative at the boundary of its support, which is always finite due to that for the particle volume. At the same time, the windowed Fourier transform suppress the LSP spectrum at all analytical points of the IAF, thus leaving peaks at non-smooth points, one of which is located on the boundary of the support and corresponds to the particle diameter $d$ (the maximum distance between two interior points) (Fig. 1). The analysis in the framework of the WKB approximation, leads to similar results [21], but the scattering coordinate should be changed on

$$\xi_m \stackrel{\text{def}}{=} \sqrt{m\xi^2 + (m-1)^2}. \tag{4}$$

Thus, the inverse transformation of the LSP (Eq. (2)) carried out along this coordinate leads to a spectrum similar to the RGD case. In particular, its shape for a homogeneous sphere does not depend on the refractive index at all.

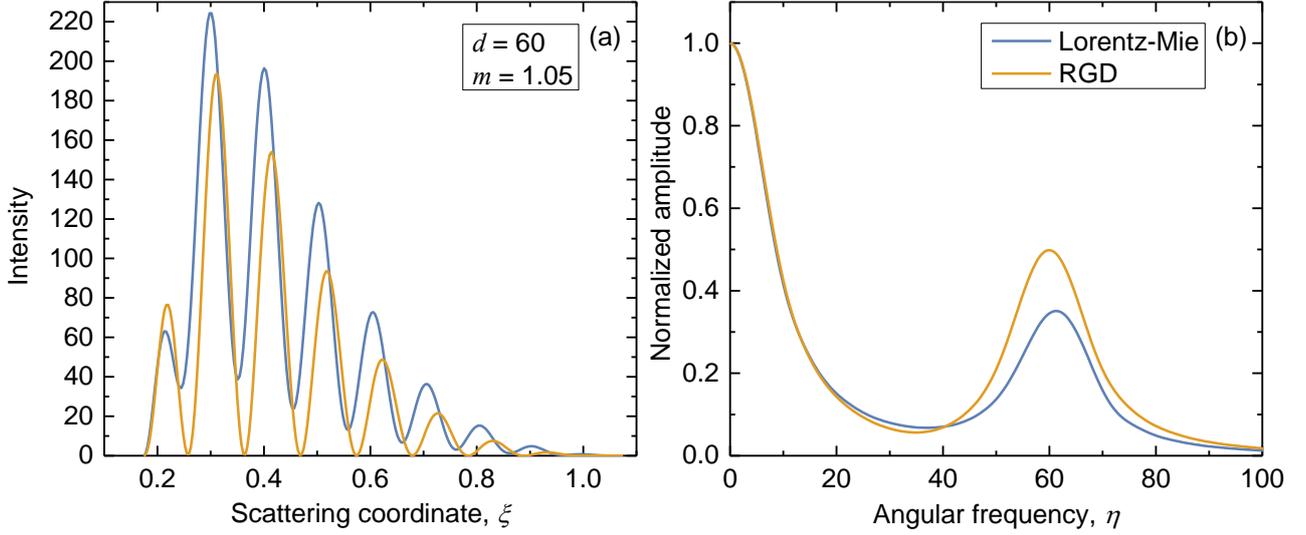

**Fig. 1.** Windowed LSP of a sphere with $d = 60$ and $m = 1.05$, calculated using the Lorenz–Mie theory (blue) and the RGD approximation (orange), as a function of the scattering coordinate $\xi$ (a), and their normalized amplitude Fourier spectrum (b).

However, without prior information about the refractive index, one cannot carry out a spectral transformation along this coordinate. Still, we can linearize it in the working range of SFC through its asymptotic expansion at infinity (Fig. 2):

$$\xi_m \approx \xi\sqrt{m} + \frac{(m-1)^2}{2\xi\sqrt{m}}. \tag{5}$$

Averaging the second term over the range $[\xi_1, \xi_2]$, we obtain the following linear approximation:

$$\xi_m \approx \sqrt{m}\,\xi + c_0, \tag{6}$$

where $c_0 \sim (m-1)^2/\sqrt{m}$.

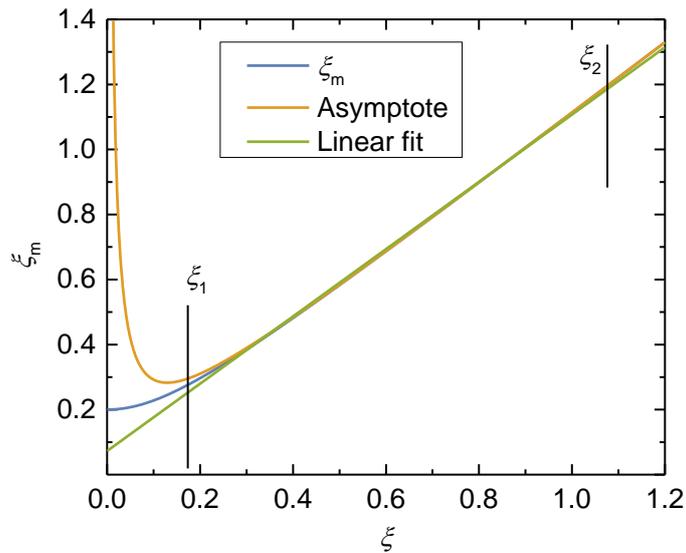

**Fig. 2.** The scattering coordinate $\xi_m$, its asymptote at infinity, and linear approximation in the range from $\xi_1$ to $\xi_2$ (corresponding to $\theta_1$ and $\theta_2$, respectively) as a function of $\xi$ for $m = 1.2$.

Thus, introducing the shifted variable $\xi = (\xi_m - c_0)/\sqrt{m}$ in Eq. (2) we obtain the result similar to the RGD spectrum, up to a well-defined phase shift:

$$F(v) = \frac{1}{\xi_2 - \xi_1} \int_{\xi_1}^{\xi_2} w(\xi)I(\xi)\exp(-iv\xi)\,d\xi \approx \exp\left(iv\frac{c_0}{\sqrt{m}}\right) F_{\text{RGD}}\left(\frac{v}{\sqrt{m}}\right), \tag{7}$$

Therefore, the transition from the RGD to the WKB spectrum boils down to the shift in the peak position $L_{\text{WKB}} = \sqrt{m}d$, accompanied by the phase shift proportional to $d(m-1)^2/\sqrt{m}$.

As can be seen from Fig. 3, the peak location and the phase of the peak have an almost linear and a close to quadratic dependence on the refractive index respectively, at least for $m < 1.1$, both in the WKB approximation and when using the rigorous Lorenz–Mie theory. Moreover, Eq. (7) correctly reproduces this trend. However, there is a noticeable difference between this simple expression and the WKB result, since it takes into account only the change of scattering coordinate (Eq. (4)), but not the change in IAF during the transition, including its anisotropy (as described in the Appendix of [21]). Interestingly, a similar behavior of phase spectrum was observed previously. For example, the phase value at peak position is known to depend on the phase-shift parameter $d(m-1)$ [19], and this dependence is closely related to the observed dependence of the positions of the minima in the light-scattering pattern on the size and refractive index [10].

Additionally, in experiments the phase is inherently a periodic quantity and can be unambiguously defined only in a limited range, like $[-\pi, \pi)$ or $[0, 2\pi)$. This, in turn, does not allow one to easily determine the refractive index in a wide range.

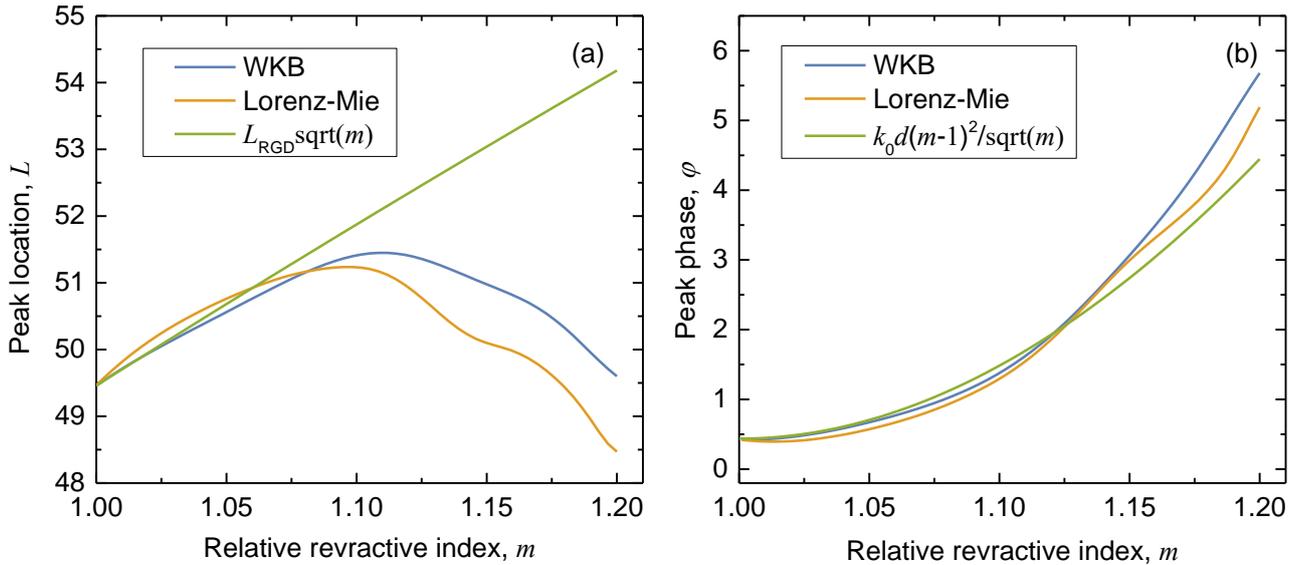

**Fig. 3.** The main peak location (a) and the phase value at that point (b) as functions of relative refractive index $m$ in the WKB approximation of a sphere and using the Lorenz–Mie theory, $d = 50$. The coefficient $k_0 = 2.22$ is obtained as a least-square fit of $k_0(m-1)^2/\sqrt{m}$ to $c_0(m)$, where the latter is obtained by analytical linear fit of Eq. (6) to Eq. (4). We chose the range of possible phase values $[0, 2\pi)$ to avoid phase discontinuities within the considered range of refractive index.

In principle, there exist various techniques for phase unwrapping [24–26]. For instance, one can follow the phase value along the spectrum from the zero frequency to the main peak. However, this does not seem robust, since the spectrum magnitude becomes small on the way, making it sensitive to all kind of uncertainties. In other words, one cannot unambiguously determine the number of full turns the corresponding value makes in the complex plane (data not shown). Therefore, in the following we circumvent this issue by considering a simpler problem, where the size and refractive index of a sphere is a priori limited to a narrow range. And these two characteristics are determined from the location and phase value of the main spectral peak.

*2.3   Inverse problem and interpolation*

To implement the characterization method, we set $d \in [44, 55]$ and $m \in [1.15, 1.22]$, which corresponds to polystyrene beads with diameter around 4 μm for incident wavelength $\lambda = 660$ nm and host medium refractive index $n_0 = 1.333$. Then we continue along the lines of Ref. [17]. We compute LSPs of spheres for a grid of values uniformly distributed in the specified ranges (the grid size is discussed below). For each of the LSPs the Fourier transform is performed using Eq. (3), from which we determined the position of the main peak and the corresponding phase value. The range of the latter was rather arbitrarily set to $[-3\pi/2, \pi/2)$ to avoid discontinuities (that is only possible in the limited range of $d$ and $m$). The resulting (inverse) mapping of the spectral parameters into particle characteristics is presented in Fig. 4.

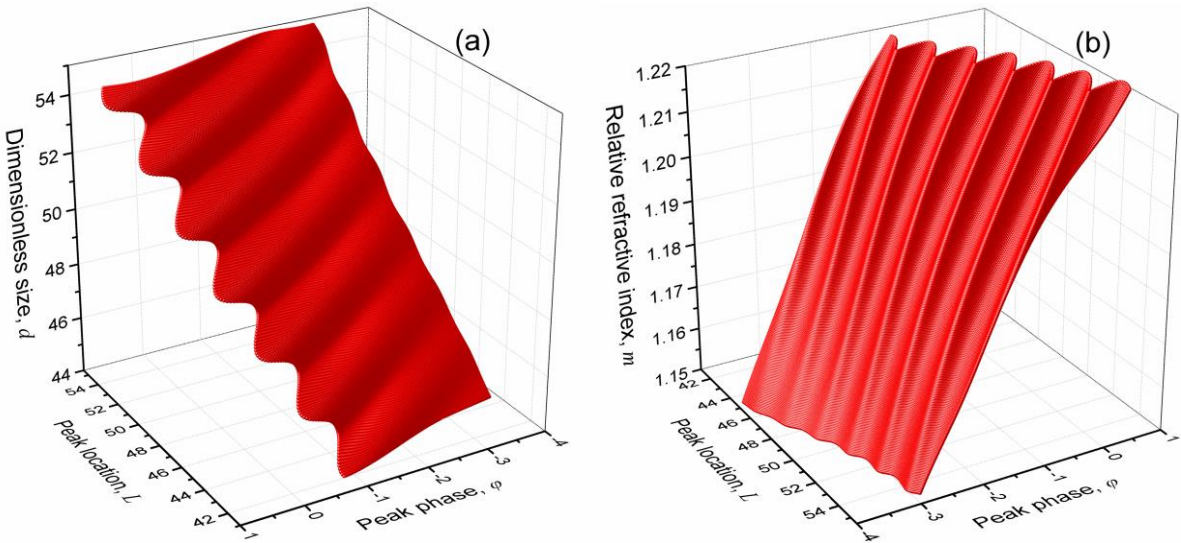

**Fig. 4.** The mapping of the spectral peak location and phase value into the size and relative refractive index of a sphere.

For routine use of this mapping, we built an interpolant by a two-stage procedure (to improve the final speed). Specifically, we interpolated the original set of points to a regular grid of spectrum parameters and used it to construct a fast interpolant. From our empirical experience, an increase in the density of points in a regular interpolant only slightly improves the accuracy of the final solution, so here and below we will use the same density of points both to calculate the direct problem (obtaining the spectrum parameters from the characteristics of the sphere) and to obtain the interpolant. Everywhere we used linear interpolation due its robustness. Importantly, the interpolant works only inside its operational range of input values, therefore, we can meaningfully discard particles with size outside of the solution region. Unfortunately, the latter is not possible for particles with outlier values of refractive index because of phase periodicity.

As can be seen in Fig. 4, the mapping has ripples, which have high derivatives and, as a result, reduce the accuracy of the solution. To quantify this issue, we show the accuracy of determining the characteristics versus the number of points used in the mapping (Fig. 5). In the following we use 256×256 grid, since it leads to maximum relative size error and absolute refractive-index error of 0.05% and 0.0001, respectively. That is more than sufficient for any potential application.

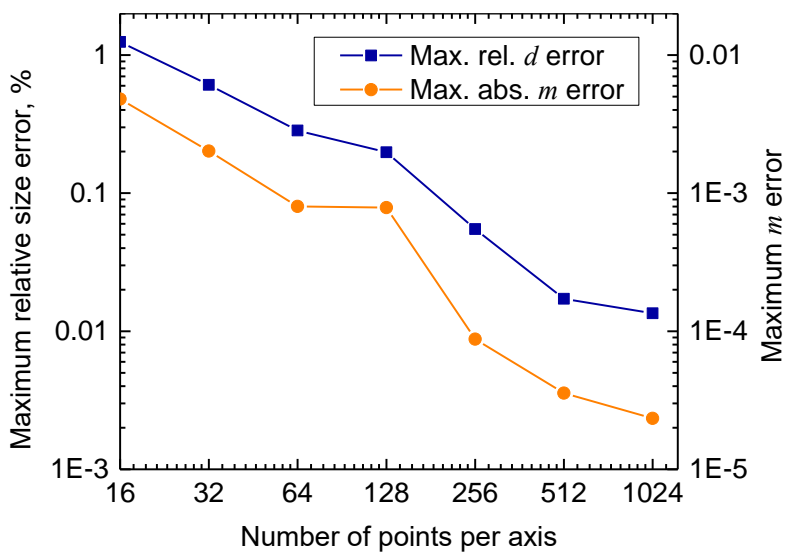

**Fig. 5.** The maximum relative size error (left axis) and absolute relative refractive index error (right axis) as functions of the number of points per axis used in construction of the direct and inverse (interpolation) mappings.

## 3 Results and discussion

### 3.1 White-noise sensitivity

In contrast to the previous similar works [17,18], here the considered range of retrieved characteristics is much smaller, so the implicit requirements for the overall accuracy are higher. Let us, first, test the method using the additive uniform white noise. Specifically, we used:

$$I_{\text{wn}}(\theta) = I(\theta) + \frac{\varepsilon}{\text{SNR}\, W(\theta)} \langle I(\theta) W(\theta) \rangle_{55°-65°}, \tag{8}$$

where $\varepsilon$ is a uniformly distributed random value in the range $[-1,1]$, SNR is a signal-to-noise ratio (used as a variable coefficient), $\langle\ \rangle_{\theta_1-\theta_2}$ is an averaging operation in range of $[\theta_1, \theta_2]$, and $W(\theta)$ is a weighting function

$$W(\theta) = \frac{1°}{\theta} \exp\left[-2\log^2\left(\frac{\theta}{54°}\right)\right], \tag{9}$$

which is an approximation of the SFC transfer function and provides an uniform experimental-noise level over the considered angular region [27]. In other words, the SNR is defined with respect to the signal amplitude in the angular range of the LSP, where it is the smallest.

In contrast to [17], we used the following mean square error (MSE) to quantify the accuracy of the results in comparison with the reference characterization method (non-linear regression):

$$\text{MSE} = \frac{1}{\theta_2 - \theta_1} \int_{\theta_1}^{\theta_2} W(\theta)^2 (I_{\text{test}}(\theta) - I_0(\theta))^2 d\theta, \tag{10}$$

where $I_{\text{test}}(\theta)$ is an LSP calculated by characteristics obtained either from the new spectral method or the reference least-square fit, $I_0(\theta)$ is either an experimental LSP or an undisturbed simulated LSP (used in the following and this sections, respectively). The MSE can also be used to quantify the simulated noise itself, using $I_{\text{wn}}$ instead of $I_{\text{test}}$. Note that the MSE depends on the overall amplitude of the signal and the number of oscillations, which, in turn, depend on the particle characteristics, especially on the size. However, this issue is alleviated by the narrow range of characteristics used in this paper. Thus, the MSE is a simple yet adequate metric for intercomparison of different methods.

We simulated 1000 LSPs with characteristics in ranges $d \in [3.8, 4.2]$ μm and $n \in [1.57, 1.61]$ (for $\lambda = 660$ nm and $n_0 = 1.333$), which are slightly smaller than the initial domain of constructed interpolant to avoid boundary effects of interpolation. In this section we switch to dimensional and absolute values for particle characteristics to avoid confusion in discussing experiments. We disturbed the data according to the Eq. (10) with SNR = 3, which is much smaller than that observed in the experiment (see next sections), i.e., the synthetic noise is much more pronounced. Then we processed it by both fitting and spectral methods and calculated MSE. Fig. 6 shows the example of noisy LSP and two simulated one, corresponding to retrieved parameters.

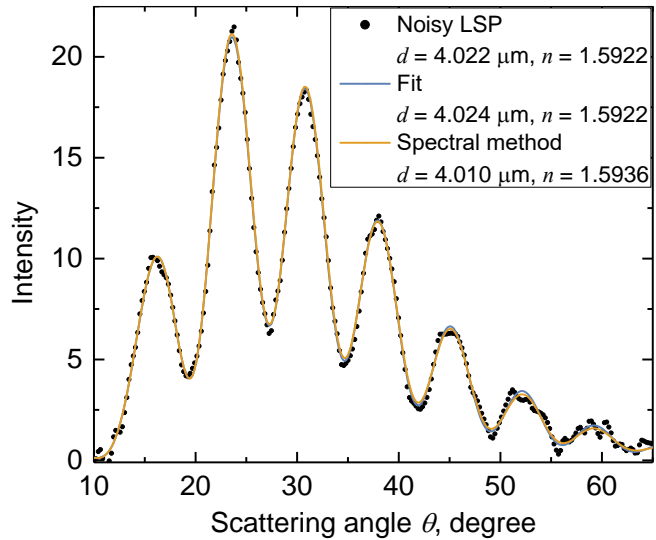

**Fig. 6.** Weighted noisy LSP (SNR = 3) of a sphere in comparison with the simulated ones. The characteristics for the latter were obtained by the spectral method and the non-linear regression and are presented in the legend. MSE and log(MSE) for the spectral method are 0.493 and −0.307, respectively, while for the regression they are 0.198 and −0.703.

A more detailed analysis of the method accuracy can be performed through the MSE distribution (Fig. 7), where we additionally show the MSE of the noisy data itself. Both characterization methods significantly decrease the MSE, i.e., the reconstructed LSPs are much closer to the undisturbed ones, then the noisy inputs. But the nonlinear regression does it more efficiently, on average 6 times smaller MSE than that for the spectral method. This is not surprising, since the reference method is based on minimizing the MSE. Anyway, we can conclude that the spectral method is resistant to white noise. Table 1 reinforces this result, demonstrating the average and maximum errors of the obtained characteristics.

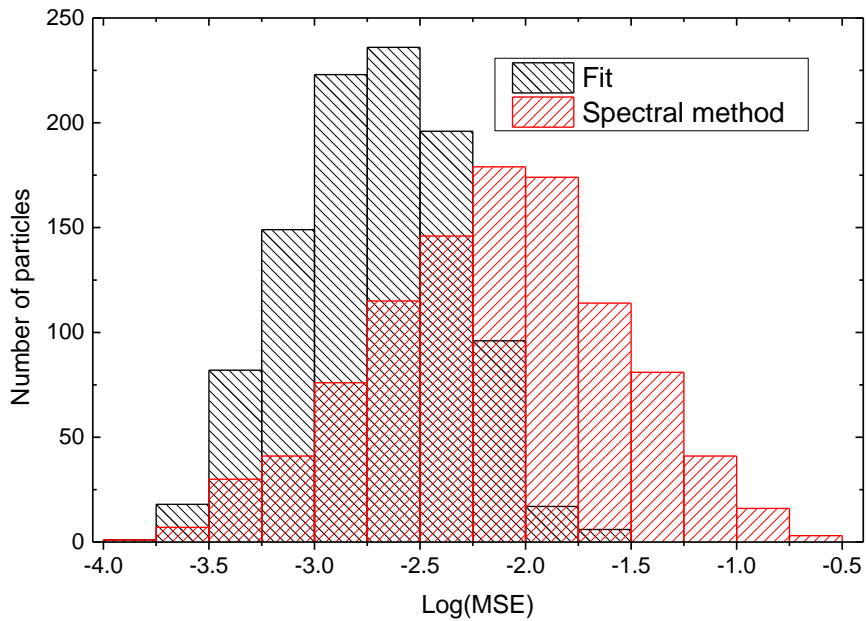

**Fig. 7.** Histogram of the decimal logarithm of MSE (see text) obtained with two methods for 1000 noisy LSPs (SNR = 3).

Table 1. Results of characterization of synthetic LSPs with SNR = 3

|  | Median relative size error, % | Maximum relative size error, % | Median absolute refractive-index error | Maximum absolute refractive-index error |
|---|---|---|---|---|
| Least-square fit | 0.04 | 0.6 | 0.0004 | 0.004 |
| Spectral method | 0.12 | 1.2 | 0.0012 | 0.008 |

### 3.2 Experimental data

The experimental verification of the developed method was performed using the SFC. We measured and processed 323 polystyrene beads with the reference non-linear regression method from [28]. Briefly, it fits an experimental LSP with simulated ones minimizing a metric similar to Eq. (10) using the global optimization algorithm DiRect. For each particle, the size, refractive index and MSE were determined using both reference and spectral methods. Additionally, the reference method estimates characterization errors, corresponding to nominal confidence of one standard deviation. Fig. 8 shows an example of an experimental signal in comparison with the simulated ones corresponding to the characteristics retrieved by both methods. The difference in retrieved characteristics is relatively small (roughly twice the standard error of the reference method) and it can be attributed to the non-trivial experimental distortions, which affects different part of LSP in varying degree. The least-square fit better reproduces the amplitudes of the largest peaks (as expected), while the spectral method is better at matching the position of maxima and minima at intermediate scattering angles (between 35° and 50°). One can even use the difference between the two methods as an estimate of the magnitude of experimental distortions.

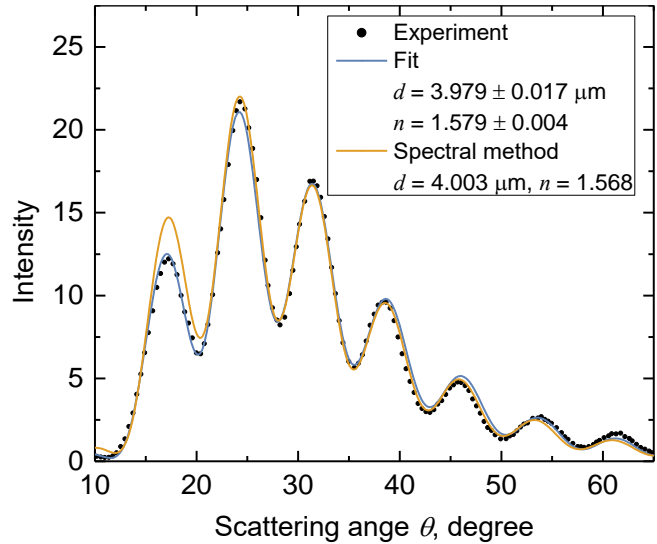

**Fig. 8.** Weighted LSP of a polystyrene sphere in comparison with the Lorenz–Mie theory. The characteristics obtained by the spectral method and the non-linear regressions are presented in the legend. MSE and log(MSE) for the spectral method are $0.493$ and $-0.307$, respectively, and for the regression are $0.198$ and $-0.703$.

The characterization results for the whole sample are presented in Fig. 9. The size distributions are almost identical for two methods, while the distribution over the refractive index obtained with the spectral method is significantly broader than the reference one, see Fig. 9(a). Knowing the literature value for the polystyrene refractive index at this wavelength, 1.584 [29], we can classify the lower part of this distribution as outliers. Still, this has only minor effect on the median value of the retrieved refractive index Table 2. Also, such shift of refractive index is reasonable for given experimental distortions. Specifically, 52% and 81% of all particles have the characteristics determined by the spectral method within 1 and 2 standard deviations, respectively, from that of the reference method. Moreover, MSE distribution of the spectral method is only slightly inferior to the reference one, as shown in Fig. 9(b). In terms of absolute values, the median differences of size and refractive index between the spectral method and the reference ones are 13 nm (0.3%) and 0.004, respectively. But this is an upper estimate of the spectral-method errors, since they are comparable to uncertainty of the reference values (Table 2).

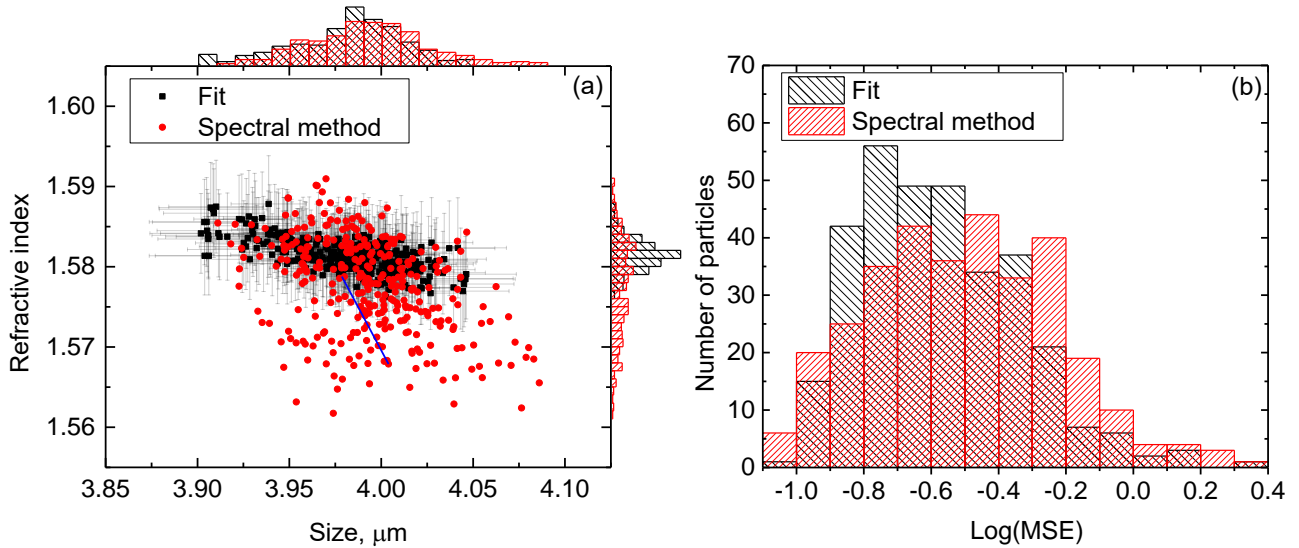

**Fig. 9.** Results of the characterization of 323 polystyrene beads by the spectral method and non-linear regression (fit). The latter also estimates the error of a single measurement, corresponding error bars (one standard deviation) are shown as in part (a). The blue line connects the retrieved characteristics of the particle shown in Fig. 8. The part (b) shows a comparison of the obtained MSE (see text) for each method in logarithmic scale.

Table 2. Median values of results presented in Fig. 9(a). The errors are based on internal estimates that are not available for the spectral method.

|  | Median size, μm | Median refractive-index | Median size error (std), μm | Median refractive-index error (std) |
|---|---|---|---|---|
| Least-square fit | 3.985 | 1.581 | 0.017 | 0.005 |
| Spectral method | 3.992 | 1.578 | - | - |

Therefore, we conclude that the developed spectral method works reliably for the considered polystyrene beads, but leads to about twice larger errors in determined refractive index as compared to the reference least-square method. At the same time the spectral method is much faster, requiring 20 ms instead of 5 s per particle on a common laptop. Moreover, the pronounced sensitivity of the spectrum phase to the experimental distortions (leading to errors in refractive-index retrievals) can be employed for precise tuning of the experimental set up.

## 4 Conclusion

We considered the transition from RGD to WKB approximations for the light-scattering by a sphere, where the change of the LSP is mostly due to the modification of the scattering coordinate $\xi$. Linearization of the modified coordinate allowed us to derive simple relations between the main parameters of the LSP spectrum, such as the position of the main peak and the phase value at this point, and the particle refractive index. These estimates, especially for the peak phase, proved to be surprisingly consistent with the numerical simulation using both the WKB approximation and the rigorous Lorenz–Mie theory.

Based on the derived relations, we developed a characterization method for single polystyrene beads using the position of the main peak in the LSP spectrum and the phase value at this point as input. To avoid phase wrapping, we limited the solution to the range of $d \in [44, 55]$ and $m \in [1.15, 1.22]$, corresponding to 4-μm polystyrene beads in water, and constructed an interpolant that maps the spectrum parameters into the particle characteristics. The number of underlying grid points was proven to be sufficient for excellent accuracy even in the presence of ripples in the direct mapping. We studied the effect of white noise on the method performance in comparison with the reference method of non-linear regression. The developed method demonstrated robustness for the noise level of $\mathrm{SNR} = 3$ with only twice larger maximum errors than that for the reference one.

We also tested the spectral method on real measured data of polystyrene beads against a reference one. Specifically, 52% and 81% of all particles have the characteristics determined by the spectral method within 1 and 2 standard deviations, respectively, from that of the reference method. The upper estimates for the median errors of the spectral method are 13 nm and 0.004 for size and refractive index, respectively. Thus, the spectral method leads to comparable accuracy (for given experimental distortions) but is about 300 times faster than non-linear regression. It is a valuable addition to the toolkit of precise characterization methods for spherical particles and can be easily adapted to any other range of size and refractive index of similar width. The future research may extend the method applicability to a much wider range by either employing phase unwrapping or combining it with a method capable of crude estimation of particle characteristics.